\documentclass[aps,prl,twocolumn,epsf]{revtex4}

\begin{document}
\title{All possible bipartite positive-operator-value measurements \\ of two-photon polarization states} 
 
\author{S. E. Ahnert} 
\author{M. C. Payne}
\affiliation{Theory of Condensed Matter Group, Cavendish Laboratory, \\
Madingley Road, Cambridge CB3 0HE, U.K.}
\begin{abstract}Here we propose an implementation of all possible Positive Operator Value Measures (POVMs) of two-photon polarization states. POVMs are the most general class of quantum measurements. Our setup requires linear optics, Bell State measurements and an entangled three-photon ancilla state, which can be prepared separately and in advance (or 'off-line'). As an example we give the detailed settings for a simultaneous measurement of all four Bell States for an arbitrary two-photon polarization state, which is impossible with linear optics alone. 

\bigskip

\end{abstract}
\maketitle 


\section{introduction}
Quantum Information Theory has been a field of increasing activity over the past two decades (see \cite{NielsenChuang} for a comprehensive overview). In the wake of this the area of Quantum Measurement has been the subject of research in recent years, in particular the theory and implementations of generalized measurement in the form of Positive Operator Value Measures (POVMs) \cite{Peres,Calsamiglia,us}.
The uses of such measurements lie in Quantum State Estimation \cite{bergou} and they are of direct practical use in Quantum Cryptography \cite{peres,ekert,brandt97,brandt99}. 

In earlier work \cite{us} we proposed a setup for performing all possible positive operator value measures of single photon polarization states. Here we report an important extension to this work, namely the implementation of all possible POVMs of {\em two-photon} polarization states. It has been shown that this cannot be achieved using linear optics alone \cite{Calsamiglia} and our setup uses measurements and an optical non-linearity to achieve its goal. As an example of such a measurement we give the details for performing a complete Bell-state measurement of a two-photon polarization state using our setup. A specific setup achieving this particular measurement with non-linear optics was implemented in the teleportation experiment of Kim {\em et al.} \cite{kim}.

In general our method works by teleporting a bipartite photon polarization state onto the path-polarization state of a single photon. Linear optics then allows for any POVM to be performed in the Hilbert space of the two degrees of freedom of this third photon. Path-polarization states were used in one of the first teleportation experiments \cite{hardypop} which included a full Bell-State measurement in this basis. In general states with entanglement in different degrees of freedom are termed {\em hyper-entangled} states, and such states have been the subject of research in recent years \cite{kwiat,cinelli,kwiat2}. 

\section{Teleportation}

Since its conception by Bennett {\em et al.} \cite{bennett} the concept of teleportation has been generalized in to continuous variables \cite{vaidman,braunstein}, $N$-dimensional quantum systems \cite{stenholm} and bipartite quantum systems \cite{lee,rigolin,cola}. In our case however we do not want to simply teleport the polarization state of two photons to another two photons elsewhere, but our aim is to transfer a bipartite polarization state such as:

\begin{equation}\label{psi}
| \Psi \rangle = a | HH \rangle + b | HV \rangle + c | VH \rangle + d | VV \rangle
\end{equation}
(with H and V denoting horizontal and vertical polarizations, respectively) onto a quantum state of a single photon in the Hilbert space of its polarization and path states, i.e.:
\begin{equation}\label{phi}
| \Phi \rangle = a | Hs_1 \rangle + b | Hs_2 \rangle + c | Vs_1 \rangle + d | Vs_2 \rangle
\end{equation} 
where $| s_1 \rangle$ and $| s_2 \rangle$ are path states of the photon.
As shown in \cite{rigolin} teleportation of a two-qubit state such as a bipartite polarization state requires the two parties Alice and Bob to share one of sixteen possible four particle states, for example the state:
\[
| g_1 \rangle = {1 \over 2} (| HHHH \rangle + | HVHV \rangle + | VHVH \rangle + | VVVV \rangle)
\]
In the standard bipartite teleportation setup, the first two photons would be with Alice and the third and fourth would be with Bob. Alice would then perform a generalized Bell measurement on her two photons of the shared state as well as the two photons of the state to be teleported. She would then transmit four classical bits to Bob, telling him which of the sixteen possible generalized Bell states she had measured, allowing him to apply the appropriate corrections to his two-particle state. 

Instead of the state $| g_1 \rangle$ we will use a state of three photons in the Hilbert space of the polarization degrees of freedom of all three photons as well as the spatial degree of freedom of one of the photons, i.e.:
\[
| \tilde{g}_1 \rangle = {1 \over 2} (| HHHs_1 \rangle + | HVHs_2 \rangle + | VHVs_1 \rangle + | VVVs_2 \rangle)
\]
where the path state identifier refers to the path state of the third photon (the polarization identifier immediately preceding it). We can create such a state by sending the state 

\[
| \gamma \rangle = {1 \over \sqrt{2}} (| H_1 H_3 \rangle + | V_1 V_3 \rangle) \otimes {1 \over \sqrt{2}} (| H_2 \rangle + | V_2 \rangle) 
\]
through a Fredkin (controlled-swap) gate such that photon 2 controls the swap of photon 3 and a vacuum state. Single-photon implementations of this gate have been suggested and discussed in \cite{milburnfredkin,fredkin2,fredkin3}. It should be noted that the state $| \tilde{g}_1 \rangle$ can be prepared separately and in advance, or 'off-line'. This means that in terms of 'on-line' resources -- those needed when an actual POVM is to be realized on a given state -- our method for implementing bipartite POVMs of two-photon polarization states requires only linear optics, Bell state measurements and the ancilla state $| \tilde{g}_1 \rangle$.  

In order to teleport a general bipartite state $| \Psi \rangle$ as given in eq. (\ref{psi}) we can rewrite the combined state of the photon pair to be measured and the three photon state $| \tilde{g}_1 \rangle$ as:
\[
| \Psi \rangle | \tilde{g}_1 \rangle = {1 \over 4} \sum_{j = 1}^{16} | g_j \rangle | \phi_j \rangle
\]
where $| g_j \rangle$ are the sixteen generalized Bell states of the polarizations of four photons (given in \cite{rigolin}) and $| \phi_j \rangle$ are sixteen variations of the state $| \Phi \rangle$ of eq. (\ref{phi}), premultiplied by all sixteen possibilities of the operator $\sigma_{pol,x}^{a_1} \sigma_{path,x}^{a_2} \sigma_{pol,z}^{a_3} \sigma_{path,z}^{a_4}$ with $a_i \in \{0,1\} \, \forall \, i$. One particularly convenient state is $| g_{16} \rangle$ in \cite{rigolin}, which in polarization notation reads:

\[
| g_{16} \rangle = {1 \over 2}(| HHVV \rangle - | HVVH \rangle - | VHHV \rangle + | VVHH \rangle)
\]

As this can be rewritten as:

\[
| g_{16} \rangle = {1 \over 2}(| H_1V_3 \rangle - | V_1H_3 \rangle) \otimes (| H_2V_4 \rangle - | V_2H_4 \rangle) 
\]
i.e. a tensor product of two singlet states of photons 1 and 3, and 2 and 4, respectively. 

By combining each of these photon pairs in a conventional beamsplitter we can, upon detecting a photon each in all four of the possible outputs, conclude that the third photon of $| \tilde{g}_1 \rangle$ has been projected into the state:
\begin{eqnarray}
| \phi_{16} \rangle &=& \sigma_{pol,x} \sigma_{path,x} \sigma_{pol,z} \sigma_{path,z} | \Phi \rangle 
\cr
&=& d | Hs_1 \rangle - c | Hs_2 \rangle - b | Vs_1 \rangle + a | Vs_2 \rangle
\end{eqnarray}
With this knowledge it is a trivial matter to create from this the state $| \Phi \rangle$ by using mirrors, polarization rotators and polarizing beamsplitters. It is on this state $| \Phi \rangle$ in the Hilbert space of path and polarization states that we can perform any four-dimensional POVM using linear optics alone. The success probability of the teleportation and therefore of our method is $1 \over 16$ if we only use $| g_{16} \rangle$. This can be raised at least to $1 \over 4$, as all of the generalized Bell states $ | g_i \rangle$ in \cite{rigolin} can be written as tensor products of Bell states for photons 1 and 3, and for photons 2 and 4, respectively. Since conventional Bell state measurements on photon polarization states are successful with probability $1 \over 2$, the probability that two such measurements are, is $1 \over 4$.

\section{implementing all four-dimensional unitary operators}

While it is very difficult to perform arbitrary unitary operations on bipartite photon polarization states, doing the same on the equally sized four-dimensional Hilbert space of the path and polarization states of a single photon is by comparison trivial. Let us consider first a 'rotator' in path state space (Fig. \ref{rotator}).

\begin{figure}

\setlength{\unitlength}{1pt}
\ifx\plotpoint\undefined\newsavebox{\plotpoint}\fi
\begin{picture}(210,210)(0,0)
\font\gnuplot=cmr10 at 10pt
\gnuplot
\put(50,50){\special{em:moveto}}
\put(70,50){\special{em:lineto}}
\put(70,70){\special{em:lineto}}
\put(50,70){\special{em:lineto}}
\put(50,50){\special{em:lineto}}
\put(70,70){\special{em:lineto}}
\put(150,150){\special{em:moveto}}
\put(170,150){\special{em:lineto}}
\put(170,170){\special{em:lineto}}
\put(150,170){\special{em:lineto}}
\put(150,150){\special{em:lineto}}
\put(170,170){\special{em:lineto}}
\put(10,60){\special{em:moveto}}
\put(160,60){\special{em:lineto}}
\put(160,210){\special{em:lineto}}
\put(60,20){\special{em:moveto}}
\put(60,160){\special{em:lineto}}
\put(210,160){\special{em:lineto}}
\put(150,50){\special{em:moveto}}
\put(170,70){\special{em:lineto}}
\put(152,48){\special{em:moveto}}
\put(172,68){\special{em:lineto}}
\put(50,150){\special{em:moveto}}
\put(70,170){\special{em:lineto}}
\put(48,152){\special{em:moveto}}
\put(68,172){\special{em:lineto}}

\put(190,150){\special{em:moveto}}
\put(195,150){\special{em:lineto}}
\put(195,170){\special{em:lineto}}
\put(190,170){\special{em:lineto}}
\put(190,150){\special{em:lineto}}
\put(100,50){\special{em:moveto}}
\put(105,50){\special{em:lineto}}
\put(105,70){\special{em:lineto}}
\put(100,70){\special{em:lineto}}
\put(100,50){\special{em:lineto}}
\put(50,100){\special{em:moveto}}
\put(70,100){\special{em:lineto}}
\put(70,105){\special{em:lineto}}
\put(50,105){\special{em:lineto}}
\put(50,100){\special{em:lineto}}
\put(50,40){\special{em:moveto}}
\put(70,40){\special{em:lineto}}
\put(70,35){\special{em:lineto}}
\put(50,35){\special{em:lineto}}
\put(50,40){\special{em:lineto}}
\put(25,55){\special{em:moveto}}
\put(30,60){\special{em:lineto}}
\put(25,65){\special{em:lineto}}
\put(155,175){\special{em:moveto}}
\put(160,180){\special{em:lineto}}
\put(165,175){\special{em:lineto}}
\put(85,55){\special{em:moveto}}
\put(90,60){\special{em:lineto}}
\put(85,65){\special{em:lineto}}
\put(55,85){\special{em:moveto}}
\put(60,90){\special{em:lineto}}
\put(65,85){\special{em:lineto}}
\put(55,25){\special{em:moveto}}
\put(60,30){\special{em:lineto}}
\put(65,25){\special{em:lineto}}
\put(155,105){\special{em:moveto}}
\put(160,110){\special{em:lineto}}
\put(165,105){\special{em:lineto}}
\put(105,155){\special{em:moveto}}
\put(110,160){\special{em:lineto}}
\put(105,165){\special{em:lineto}}
\put(175,155){\special{em:moveto}}
\put(180,160){\special{em:lineto}}
\put(175,165){\special{em:lineto}}
\put(80,38){\makebox(0,0){$-{\pi \over 2}$}}
\put(192,140){\makebox(0,0){$\pi \over 2$}}
\put(102,40){\makebox(0,0){$\alpha$}}
\put(40,102){\makebox(0,0){$\alpha$}}
\put(75,25){\makebox(0,0){$| s_2 \rangle$}}
\put(35,75){\makebox(0,0){$| s_1 \rangle$}}
\put(205,170){\makebox(0,0){$| s_2 \rangle$}}
\put(170,205){\makebox(0,0){$| s_1 \rangle$}}
\put(35,35){\makebox(0,0){$| \Phi \rangle$}}
\put(85,80){\makebox(0,0){PBS}}
\put(135,175){\makebox(0,0){PBS}}
\end{picture}

\caption{The rotator in path state space utilizing two polarizing beamsplitters (PBS) and four polarization rotators ($\alpha$, $\pi \over 2$, $-{\pi \over 2}$). A single photon path-polarization state $| \Phi \rangle$ incident on the two entrances at the bottom left will be rotated in the path-state basis $\{| s_1 \rangle, | s_2 \rangle\}$ by angle $\alpha$. }\label{rotator}
\end{figure}

The superposition path state incident on the entrances of the polarizing beamsplitter at the bottom left of Fig. \ref{rotator} evolves as follows:
\begin{eqnarray}
| \Phi \rangle &=& a | Hs_1 \rangle + b | Vs_1 \rangle + c | Hs_2 \rangle + d | Vs_2 \rangle \cr\cr
&\rightarrow& (a \cos \alpha + c \sin \alpha) | Hs_1 \rangle + (d \sin \alpha + b \cos \alpha) | Vs_1 \rangle \cr
&+&  (- a \sin \alpha + c \cos \alpha)  | Hs_2 \rangle + (d \cos \alpha - b \sin \alpha) | Vs_2 \rangle \cr
\end{eqnarray}
or, in matrix notation:
\begin{eqnarray}
\left(
\begin{array}{c}
a \cr b \cr c \cr d
\end{array}
\right)
\rightarrow
\left(
\begin{array}{cccc}
\cos \alpha & 0 & \sin \alpha & 0 \cr
0 & \cos \alpha & 0 & \sin \alpha \cr
- \sin \alpha & 0 & \cos \alpha & 0 \cr
0 & - \sin \alpha & 0 & \cos \alpha \cr
\end{array}
\right)
\left(
\begin{array}{c}
a \cr b \cr c \cr d 
\end{array}
\right)
\cr\cr\cr
= 
\left(
\begin{array}{cc}
{\bf 1} \cos \alpha  & {\bf 1} \sin \alpha  \cr
- {\bf 1} \sin \alpha  & {\bf 1} \cos \alpha  \cr
\end{array}
\right) | \Phi \rangle
=
R_{path}(\alpha) | \Phi \rangle
\end{eqnarray}
where ${\bf 1}$ denotes the $2 \times 2$ unit matrix. Thus the setup in Fig. \ref{rotator} performs rotations in the path state basis.
To turn this into a general unitary operation on the path state space we introduce phase shifters at both entrances and both exits, obtaining the matrix:
\begin{eqnarray}
\left(
\begin{array}{cc}
e^{i {\zeta+\xi \over 2}} {\bf 1} & 0 \cr
0 & e^{-i {\zeta+\xi \over 2}} {\bf 1} \cr
\end{array}
\right) R_{path}(\alpha) 
\left(
\begin{array}{cc}
e^{i {\zeta-\xi \over 2}} {\bf 1} & 0 \cr
0 & e^{-i {\zeta-\xi \over 2}} {\bf 1} \cr
\end{array}
\right)
\cr\cr
=\left(
\begin{array}{cc}
e^{i \zeta} {\bf 1} \cos \alpha &  e^{i \xi} {\bf 1} \sin \alpha \cr
-  e^{-i \xi} {\bf 1} \sin \alpha &  e^{-i \zeta} {\bf 1} \cos \alpha\cr
\end{array}
\right)
\cr\cr
=
U_{path}(\alpha, \zeta, \xi) | \Phi \rangle
\end{eqnarray}
If we now consider unitary operations in each of the path states before ($V_1,V_2$) and after ($U_1,U_2$) the setup of Fig. \ref{rotator}, we arrive at a general rotation matrix $\tilde{U}$:
\begin{eqnarray}
\left(\begin{array}{cc}
U_1 & 0 \cr
0 & U_2 \cr
\end{array}\right)
U_{path}(\alpha)
\left(\begin{array}{cc}
V_1 & 0 \cr
0 & V_2 \cr
\end{array}\right)
&=&
\cr\cr
\left(
\begin{array}{cc}
 U_1 V_1 e^{i \zeta} \cos \alpha  &  U_1 V_2 e^{i \xi}  \sin \alpha  \cr
- U_2 V_1 e^{-i \xi} \sin \alpha  &  U_2 V_2 e^{-i \zeta} \cos \alpha \cr
\end{array}
\right) &=& \tilde{U} 
\end{eqnarray}
This matrix represents SU(4), as we can write any four dimensional state vector in Hilbert space as:
\[
| s_\theta \rangle
= \left( 
\begin{array}{l}
\cos (\theta_1) \cos (\theta_2) e^{i(\theta_3+\theta_4)}\cr
\cos (\theta_1) \sin (\theta_2) e^{i(\theta_3-\theta_4)}\cr
\sin (\theta_1) \cos (\theta_5) e^{i(-\theta_3+\theta_6)}\cr
\sin (\theta_1) \sin (\theta_5) e^{i(-\theta_3-\theta_6)}\cr
\end{array}
\right)
\]
and thus to perform a unitary transformation from this vector to another arbitrary state vector $| s_{\theta'} \rangle$, we choose $\alpha = \theta'_1 - \theta_1$, $\zeta = \theta'_3 - \theta_3$, $\xi = \theta'_3 + \theta_3$, $U_1 = U(-\theta_2, -\theta_4, \theta_4)$, $U_2 = U(-\theta_5, -\theta_6, \theta_6)$, $V_1 = U(\theta'_2, \theta'_4, -\theta'_4)$ and $V_2 = U(\theta'_5, \theta'_6, -\theta'_6)$, where:
\[
U(\theta_i, \theta_j, \theta_k)
=
\left(
\begin{array}{cc}
e^{i \theta_j} \cos \theta_i  & - e^{i \theta_k}  \sin \theta_i  \cr
e^{-i \theta_k} \sin \theta_i  &  e^{-i \theta_j} \cos \theta_i \cr
\end{array}
\right)
\]
which is a general $2 \times 2$ unitary matrix.


\section{the single photon povm module}

The positive operator value measure (POVM) is the most general formulation of quantum measurement \cite{bibkraus}. Mathematically it corresponds to a positive-definite partition of unity in the space of operators on a given Hilbert space. A POVM is given by a set of positive definite Hermitian operators $\{F_i\}$, which in turn can be expressed in terms a set of so-called Kraus operators $\{M_i\}$, such that $F_i = M^\dagger_i M_i$ and for a POVM with $n$ operators $\sum_{i=1}^n M_i^\dagger M_i = \sum_{i=1}^n F_i = I$, where $I$ is the unit matrix.
 After a POVM measurement is performed on a quantum state represented by a density matrix $\rho$, the state becomes $\rho' = {M_i \rho M_i^\dagger \over {\rm tr}(M_i \rho M_i^\dagger)}$ with probability $p_i = {\rm tr}(M_i \rho M_i^\dagger)$.

\begin{figure}

\setlength{\unitlength}{0.8pt}
\ifx\plotpoint\undefined\newsavebox{\plotpoint}\fi
\begin{picture}(230,270)(0,0)
\font\gnuplot=cmr10 at 10pt
\gnuplot
\put(50,50){\special{em:moveto}}
\put(70,50){\special{em:lineto}}
\put(70,70){\special{em:lineto}}
\put(50,70){\special{em:lineto}}
\put(50,50){\special{em:lineto}}
\put(70,70){\special{em:lineto}}
\put(190,190){\special{em:moveto}}
\put(210,190){\special{em:lineto}}
\put(210,210){\special{em:lineto}}
\put(190,210){\special{em:lineto}}
\put(190,190){\special{em:lineto}}
\put(210,210){\special{em:lineto}}
\put(150,150){\special{em:moveto}}
\put(170,150){\special{em:lineto}}
\put(170,170){\special{em:lineto}}
\put(150,170){\special{em:lineto}}
\put(150,150){\special{em:lineto}}
\put(170,170){\special{em:lineto}}
\put(150,50){\special{em:moveto}}
\put(170,50){\special{em:lineto}}
\put(170,70){\special{em:lineto}}
\put(150,70){\special{em:lineto}}
\put(150,50){\special{em:lineto}}
\put(170,70){\special{em:lineto}}
\put(50,150){\special{em:moveto}}
\put(70,150){\special{em:lineto}}
\put(70,170){\special{em:lineto}}
\put(50,170){\special{em:lineto}}
\put(50,150){\special{em:lineto}}
\put(70,170){\special{em:lineto}}
\put(10,60){\special{em:moveto}}
\put(200,60){\special{em:lineto}}
\put(200,250){\special{em:lineto}}
\put(60,60){\special{em:moveto}}
\put(60,200){\special{em:lineto}}
\put(200,200){\special{em:lineto}}
\put(60,160){\special{em:moveto}}
\put(160,160){\special{em:lineto}}
\put(160,60){\special{em:moveto}}
\put(160,250){\special{em:lineto}}
\put(190,50){\special{em:moveto}}
\put(210,70){\special{em:lineto}}
\put(192,48){\special{em:moveto}}
\put(212,68){\special{em:lineto}}
\put(50,190){\special{em:moveto}}
\put(70,210){\special{em:lineto}}
\put(48,192){\special{em:moveto}}
\put(68,212){\special{em:lineto}}

\put(135,50){\special{em:moveto}}
\put(140,50){\special{em:lineto}}
\put(140,70){\special{em:lineto}}
\put(135,70){\special{em:lineto}}
\put(135,50){\special{em:lineto}}
\put(100,50){\special{em:moveto}}
\put(105,50){\special{em:lineto}}
\put(105,70){\special{em:lineto}}
\put(100,70){\special{em:lineto}}
\put(100,50){\special{em:lineto}}
\put(80,190){\special{em:moveto}}
\put(75,190){\special{em:lineto}}
\put(75,210){\special{em:lineto}}
\put(80,210){\special{em:lineto}}
\put(80,190){\special{em:lineto}}
\put(150,80){\special{em:moveto}}
\put(170,80){\special{em:lineto}}
\put(170,85){\special{em:lineto}}
\put(150,85){\special{em:lineto}}
\put(150,80){\special{em:lineto}}
\put(190,130){\special{em:moveto}}
\put(210,130){\special{em:lineto}}
\put(210,135){\special{em:lineto}}
\put(190,135){\special{em:lineto}}
\put(190,130){\special{em:lineto}}
\put(130,190){\special{em:moveto}}
\put(130,210){\special{em:lineto}}
\put(135,210){\special{em:lineto}}
\put(135,190){\special{em:lineto}}
\put(130,190){\special{em:lineto}}
\put(50,100){\special{em:moveto}}
\put(70,100){\special{em:lineto}}
\put(70,105){\special{em:lineto}}
\put(50,105){\special{em:lineto}}
\put(50,100){\special{em:lineto}}
\put(25,55){\special{em:moveto}}
\put(30,60){\special{em:lineto}}
\put(25,65){\special{em:lineto}}
\put(155,185){\special{em:moveto}}
\put(160,190){\special{em:lineto}}
\put(165,185){\special{em:lineto}}
\put(195,215){\special{em:moveto}}
\put(200,220){\special{em:lineto}}
\put(205,215){\special{em:lineto}}
\put(85,55){\special{em:moveto}}
\put(90,60){\special{em:lineto}}
\put(85,65){\special{em:lineto}}
\put(155,105){\special{em:moveto}}
\put(160,110){\special{em:lineto}}
\put(165,105){\special{em:lineto}}
\put(105,155){\special{em:moveto}}
\put(110,160){\special{em:lineto}}
\put(105,165){\special{em:lineto}}
\put(55,85){\special{em:moveto}}
\put(60,90){\special{em:lineto}}
\put(65,85){\special{em:lineto}}
\put(105,195){\special{em:moveto}}
\put(110,200){\special{em:lineto}}
\put(105,205){\special{em:lineto}}
\put(180,55){\special{em:moveto}}
\put(185,60){\special{em:lineto}}
\put(180,65){\special{em:lineto}}
\put(55,180){\special{em:moveto}}
\put(60,185){\special{em:lineto}}
\put(65,180){\special{em:lineto}}
\put(195,105){\special{em:moveto}}
\put(200,110){\special{em:lineto}}
\put(205,105){\special{em:lineto}}
\put(190,250){\special{em:moveto}}
\put(195,250){\special{em:lineto}}
\put(200,250){\special{em:moveto}}
\put(205,250){\special{em:lineto}}
\put(210,250){\special{em:moveto}}
\put(210,255){\special{em:lineto}}
\put(210,260){\special{em:moveto}}
\put(210,265){\special{em:lineto}}
\put(210,270){\special{em:moveto}}
\put(205,270){\special{em:lineto}}
\put(200,270){\special{em:moveto}}
\put(195,270){\special{em:lineto}}
\put(190,270){\special{em:moveto}}
\put(190,265){\special{em:lineto}}
\put(190,260){\special{em:moveto}}
\put(190,255){\special{em:lineto}}
\put(150,250){\special{em:moveto}}
\put(155,250){\special{em:lineto}}
\put(160,250){\special{em:moveto}}
\put(165,250){\special{em:lineto}}
\put(170,250){\special{em:moveto}}
\put(170,255){\special{em:lineto}}
\put(170,260){\special{em:moveto}}
\put(170,265){\special{em:lineto}}
\put(170,270){\special{em:moveto}}
\put(165,270){\special{em:lineto}}
\put(160,270){\special{em:moveto}}
\put(155,270){\special{em:lineto}}
\put(150,270){\special{em:moveto}}
\put(150,265){\special{em:lineto}}
\put(150,260){\special{em:moveto}}
\put(150,255){\special{em:lineto}}
\put(150,120){\special{em:moveto}}
\put(170,120){\special{em:lineto}}
\put(170,125){\special{em:lineto}}
\put(150,125){\special{em:lineto}}
\put(150,120){\special{em:lineto}}
\put(190,160){\special{em:moveto}}
\put(210,160){\special{em:lineto}}
\put(210,165){\special{em:lineto}}
\put(190,165){\special{em:lineto}}
\put(190,160){\special{em:lineto}}
\put(150,220){\special{em:moveto}}
\put(170,220){\special{em:lineto}}
\put(170,225){\special{em:lineto}}
\put(150,225){\special{em:lineto}}
\put(150,220){\special{em:lineto}}
\put(190,235){\special{em:moveto}}
\put(210,235){\special{em:lineto}}
\put(210,240){\special{em:lineto}}
\put(190,240){\special{em:lineto}}
\put(190,235){\special{em:lineto}}
\put(45,50){\special{em:moveto}}
\put(40,50){\special{em:lineto}}
\put(40,70){\special{em:lineto}}
\put(45,70){\special{em:lineto}}
\put(45,50){\special{em:lineto}}

\put(102,40){\makebox(0,0){$\theta$}}
\put(132,220){\makebox(0,0){$\pi$}}
\put(40,102){\makebox(0,0){$\phi$}}
\put(138,40){\makebox(0,0){$\pi \over 2$}}
\put(78,220){\makebox(0,0){$\pi \over 2$}}
\put(180,82){\makebox(0,0){$-\pi \over 2$}}
\put(220,132){\makebox(0,0){$\pi$}}
\put(220,200){\makebox(0,0){$P_2$}}
\put(180,160){\makebox(0,0){$P_1$}}
\put(215,220){\makebox(0,0){$| p_2 \rangle$}}
\put(175,190){\makebox(0,0){$| p_1 \rangle$}}
\put(75,85){\makebox(0,0){$| s_2 \rangle$}}
\put(80,45){\makebox(0,0){$| s_1 \rangle$}}
\put(180,45){\makebox(0,0){$| t_1 \rangle$}}
\put(145,95){\makebox(0,0){$| t_2 \rangle$}}
\put(80,145){\makebox(0,0){$| t_3 \rangle$}}
\put(45,180){\makebox(0,0){$| t_4 \rangle$}}
\put(140,122){\makebox(0,0){$e^{i\beta}$}}
\put(220,162){\makebox(0,0){$e^{i\gamma}$}}
\put(42,40){\makebox(0,0){$U^s$}}
\put(180,222){\makebox(0,0){$V^s_1$}}
\put(220,237){\makebox(0,0){$V^s_2$}}
\put(15,50){\makebox(0,0){$| \psi \rangle$}}
\put(160,260){\makebox(0,0){$E_1$}}
\put(200,260){\makebox(0,0){$E_2$}}
\end{picture}

\caption{The module implementing any single photon 2-operator POVM. The photon enters in state $| \psi \rangle$ at the bottom left corner and exits either at $E_1$ or $E_2$, where it can be detected. All beamsplitters are polarizing beamsplitters with the same polarization basis and transmit photons in the $| H \rangle$ state, while reflecting photons in the $| V \rangle$ state. The angles $\theta$, $\phi$, $\pi \over 2$ and $\pi$ of the polarization rotators are measured relative to this basis. $U^s$, $V^s_1$ and $V^s_2$ are unitary operators, and $e^{i \beta}$ and $e^{i \gamma}$ signify phase shifters.}\label{module}
\end{figure}
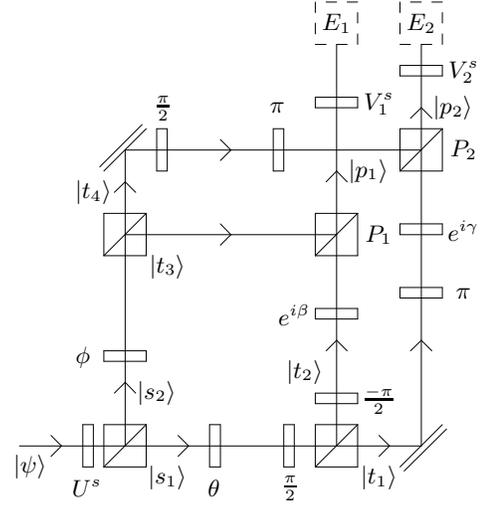

In previous work \cite{us} we introduced an implementation of all possible single photon POVMs. Using the module depicted in Fig. \ref{module} one can achieve any bipartition of unity in the form of an arbitrary pair of Kraus operators on the Hilbert space of the single photon polarization state. Using this module iteratively it is then possible to implement any set of Kraus operators.

As discussed in \cite{us} this module implements the general two-operator POVM given by the operators $F_1$ and $F_2$:
\begin{equation}
F_1 = M_1^{\dagger} M_1 = U^{s \dagger} D_1^2 U^s
\end{equation}
where $D_1^2 \equiv D_1^{\dagger} V_1^{s \dagger} V^s_1 D_1 = D_1^{\dagger} D_1$, and
\begin{equation}
D_1=
\left(
\begin{array}{cc}
e^{i\beta} \cos \theta & 0 \cr 0 & \cos \phi 
\end{array}
\right)
\end{equation}
so that:
\begin{eqnarray}
F_2 = I - F_1 = I - U^{s \dagger} D_1^2 U^s = U^{s \dagger} U^s - U^{s \dagger} D_1^2 U^s 
\cr 
= U^{s \dagger} (I - D_1^2) U^s = U^{s \dagger} D_2^2 U^s
\end{eqnarray}
where $D_2^2 \equiv D_2^{\dagger} V_2^{s \dagger} V^s_2 D_2 = D_2^{\dagger} D_2$, and
\begin{equation}
D_2=
\left(
\begin{array}{cc}
e^{i\gamma} \sin \theta & 0 \cr 0 & \sin \phi 
\end{array}
\right)
\end{equation}
Placing one of these single-photon modules with $U^s = V^s_1 = V^s_2 = I$ into each of the two paths associated with the states $| s_1 \rangle$ and $| s_2 \rangle$, and combining this with the SU(4) prerotation matrix $\tilde{U}$ gives rise to a general set of four four-dimensional POVM operators:
\begin{eqnarray}
F_1 = \tilde{U}^{\dagger}\left(\begin{array}{cccc}
D_1^2(\theta_1,\phi_1) & {\bf 0} \cr
{\bf 0} & {\bf 0}\cr
\end{array}\right)
\tilde{U}
\cr\cr
F_2 = \tilde{U}^{\dagger}\left(\begin{array}{cccc}
D_2^2(\theta_1,\phi_1) & {\bf 0} \cr
{\bf 0} & {\bf 0}\cr
\end{array}\right)
\tilde{U}
\cr\cr
F_3 = \tilde{U}^{\dagger}\left(\begin{array}{cccc}
{\bf 0} & {\bf 0}\cr
{\bf 0} & D_1^2(\theta_2,\phi_2) \cr
\end{array}\right)
\tilde{U}
\cr\cr
F_4 = \tilde{U}^{\dagger}\left(\begin{array}{cccc}
{\bf 0} & {\bf 0}\cr
{\bf 0} & D_2^2(\theta_2,\phi_2) \cr
\end{array}\right)
\tilde{U}
\end{eqnarray}
This in fact is an implementation of the most general four-dimensional bipartition of unity, into the operators $F_1+F_3$ and $F_2+F_4$ (or we could alternatively choose $F_1 + F_4$ and $F_2 + F_3$). In order to implement the most general pair of POVM operators this is enough. For the most general pair of Kraus operators we would have to apply SU(4) rotations of the type $\tilde{U}$ to the pairs of exits 1 and 3, and 2 and 4 respectively, in order to implement what in the single particle case were the operators $V_1^s$ and $V_2^s$.

The generalization of the two-photon setup is straightforward, similar to that for one photon as seen in \cite{us}. To form a further general partition of one of the operators $F_1+F_3$ or $F_2+F_4$, a SU(4) rotation is applied to the two exits of the given operator, followed by an additional two single photon POVM modules.

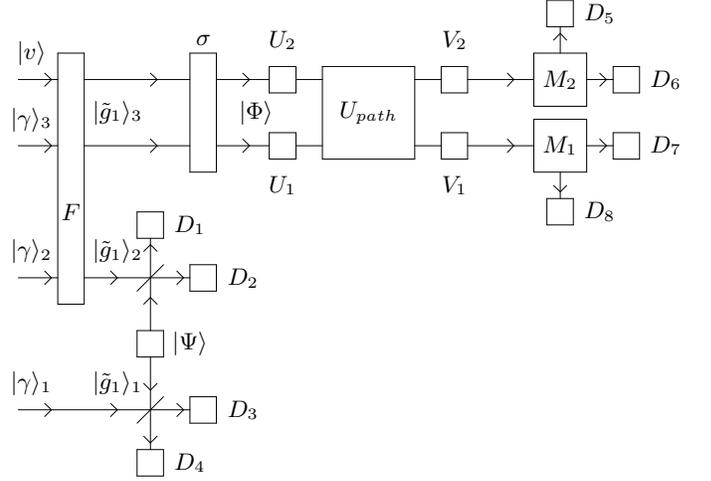
\begin{figure}

\setlength{\unitlength}{0.5pt}
\ifx\plotpoint\undefined\newsavebox{\plotpoint}\fi
\begin{picture}(550,350)(50,-20)
\font\gnuplot=cmr10 at 10pt
\gnuplot
\put(50,50){\special{em:moveto}}
\put(180,50){\special{em:lineto}}
\put(50,150){\special{em:moveto}}
\put(80,150){\special{em:lineto}}
\put(100,150){\special{em:moveto}}
\put(180,150){\special{em:lineto}}
\put(50,250){\special{em:moveto}}
\put(80,250){\special{em:lineto}}
\put(100,250){\special{em:moveto}}
\put(180,250){\special{em:lineto}}
\put(200,250){\special{em:moveto}}
\put(240,250){\special{em:lineto}}
\put(260,250){\special{em:moveto}}
\put(280,250){\special{em:lineto}}
\put(350,250){\special{em:moveto}}
\put(370,250){\special{em:lineto}}
\put(390,250){\special{em:moveto}}
\put(440,250){\special{em:lineto}}
\put(480,250){\special{em:moveto}}
\put(500,250){\special{em:lineto}}
\put(50,300){\special{em:moveto}}
\put(80,300){\special{em:lineto}}
\put(100,300){\special{em:moveto}}
\put(180,300){\special{em:lineto}}
\put(200,300){\special{em:moveto}}
\put(240,300){\special{em:lineto}}
\put(260,300){\special{em:moveto}}
\put(280,300){\special{em:lineto}}
\put(350,300){\special{em:moveto}}
\put(370,300){\special{em:lineto}}
\put(390,300){\special{em:moveto}}
\put(440,300){\special{em:lineto}}
\put(480,300){\special{em:moveto}}
\put(500,300){\special{em:lineto}}
\put(80,130){\special{em:moveto}}
\put(100,130){\special{em:lineto}}
\put(100,320){\special{em:lineto}}
\put(80,320){\special{em:lineto}}
\put(80,130){\special{em:lineto}}
\put(180,230){\special{em:moveto}}
\put(200,230){\special{em:lineto}}
\put(200,320){\special{em:lineto}}
\put(180,320){\special{em:lineto}}
\put(180,230){\special{em:lineto}}
\put(140,140){\special{em:moveto}}
\put(160,160){\special{em:lineto}}
\put(140,40){\special{em:moveto}}
\put(160,60){\special{em:lineto}}
\put(150,20){\special{em:moveto}}
\put(150,90){\special{em:lineto}}
\put(150,110){\special{em:moveto}}
\put(150,180){\special{em:lineto}}
\put(240,240){\special{em:moveto}}
\put(240,260){\special{em:lineto}}
\put(260,260){\special{em:lineto}}
\put(260,240){\special{em:lineto}}
\put(240,240){\special{em:lineto}}
\put(240,290){\special{em:moveto}}
\put(240,310){\special{em:lineto}}
\put(260,310){\special{em:lineto}}
\put(260,290){\special{em:lineto}}
\put(240,290){\special{em:lineto}}
\put(280,240){\special{em:moveto}}
\put(280,310){\special{em:lineto}}
\put(350,310){\special{em:lineto}}
\put(350,240){\special{em:lineto}}
\put(280,240){\special{em:lineto}}
\put(440,230){\special{em:moveto}}
\put(440,270){\special{em:lineto}}
\put(480,270){\special{em:lineto}}
\put(480,230){\special{em:lineto}}
\put(440,230){\special{em:lineto}}
\put(440,280){\special{em:moveto}}
\put(440,320){\special{em:lineto}}
\put(480,320){\special{em:lineto}}
\put(480,280){\special{em:lineto}}
\put(440,280){\special{em:lineto}}
\put(370,240){\special{em:moveto}}
\put(370,260){\special{em:lineto}}
\put(390,260){\special{em:lineto}}
\put(390,240){\special{em:lineto}}
\put(370,240){\special{em:lineto}}
\put(370,290){\special{em:moveto}}
\put(370,310){\special{em:lineto}}
\put(390,310){\special{em:lineto}}
\put(390,290){\special{em:lineto}}
\put(370,290){\special{em:lineto}}
\put(500,240){\special{em:moveto}}
\put(500,260){\special{em:lineto}}
\put(520,260){\special{em:lineto}}
\put(520,240){\special{em:lineto}}
\put(500,240){\special{em:lineto}}
\put(500,290){\special{em:moveto}}
\put(500,310){\special{em:lineto}}
\put(520,310){\special{em:lineto}}
\put(520,290){\special{em:lineto}}
\put(500,290){\special{em:lineto}}
\put(450,190){\special{em:moveto}}
\put(450,210){\special{em:lineto}}
\put(470,210){\special{em:lineto}}
\put(470,190){\special{em:lineto}}
\put(450,190){\special{em:lineto}}
\put(450,340){\special{em:moveto}}
\put(450,360){\special{em:lineto}}
\put(470,360){\special{em:lineto}}
\put(470,340){\special{em:lineto}}
\put(450,340){\special{em:lineto}}
\put(140,180){\special{em:moveto}}
\put(140,200){\special{em:lineto}}
\put(160,200){\special{em:lineto}}
\put(160,180){\special{em:lineto}}
\put(140,180){\special{em:lineto}}
\put(140,20){\special{em:moveto}}
\put(140,0){\special{em:lineto}}
\put(160,0){\special{em:lineto}}
\put(160,20){\special{em:lineto}}
\put(140,20){\special{em:lineto}}
\put(140,90){\special{em:moveto}}
\put(140,110){\special{em:lineto}}
\put(160,110){\special{em:lineto}}
\put(160,90){\special{em:lineto}}
\put(140,90){\special{em:lineto}}
\put(180,140){\special{em:moveto}}
\put(180,160){\special{em:lineto}}
\put(200,160){\special{em:lineto}}
\put(200,140){\special{em:lineto}}
\put(180,140){\special{em:lineto}}
\put(180,40){\special{em:moveto}}
\put(180,60){\special{em:lineto}}
\put(200,60){\special{em:lineto}}
\put(200,40){\special{em:lineto}}
\put(180,40){\special{em:lineto}}
\put(460,320){\special{em:moveto}}
\put(460,340){\special{em:lineto}}
\put(460,230){\special{em:moveto}}
\put(460,210){\special{em:lineto}}

\put(70,45){\special{em:moveto}}
\put(75,50){\special{em:lineto}}
\put(70,55){\special{em:lineto}}
\put(70,145){\special{em:moveto}}
\put(75,150){\special{em:lineto}}
\put(70,155){\special{em:lineto}}
\put(70,245){\special{em:moveto}}
\put(75,250){\special{em:lineto}}
\put(70,255){\special{em:lineto}}
\put(70,295){\special{em:moveto}}
\put(75,300){\special{em:lineto}}
\put(70,305){\special{em:lineto}}
\put(120,145){\special{em:moveto}}
\put(125,150){\special{em:lineto}}
\put(120,155){\special{em:lineto}}
\put(120,45){\special{em:moveto}}
\put(125,50){\special{em:lineto}}
\put(120,55){\special{em:lineto}}
\put(170,145){\special{em:moveto}}
\put(175,150){\special{em:lineto}}
\put(170,155){\special{em:lineto}}
\put(170,45){\special{em:moveto}}
\put(175,50){\special{em:lineto}}
\put(170,55){\special{em:lineto}}
\put(150,245){\special{em:moveto}}
\put(155,250){\special{em:lineto}}
\put(150,255){\special{em:lineto}}
\put(150,295){\special{em:moveto}}
\put(155,300){\special{em:lineto}}
\put(150,305){\special{em:lineto}}

\put(220,245){\special{em:moveto}}
\put(225,250){\special{em:lineto}}
\put(220,255){\special{em:lineto}}
\put(220,295){\special{em:moveto}}
\put(225,300){\special{em:lineto}}
\put(220,305){\special{em:lineto}}
\put(420,245){\special{em:moveto}}
\put(425,250){\special{em:lineto}}
\put(420,255){\special{em:lineto}}
\put(420,295){\special{em:moveto}}
\put(425,300){\special{em:lineto}}
\put(420,305){\special{em:lineto}}
\put(490,245){\special{em:moveto}}
\put(495,250){\special{em:lineto}}
\put(490,255){\special{em:lineto}}
\put(490,295){\special{em:moveto}}
\put(495,300){\special{em:lineto}}
\put(490,305){\special{em:lineto}}

\put(145,30){\special{em:moveto}}
\put(150,25){\special{em:lineto}}
\put(155,30){\special{em:lineto}}
\put(145,70){\special{em:moveto}}
\put(150,65){\special{em:lineto}}
\put(155,70){\special{em:lineto}}
\put(145,130){\special{em:moveto}}
\put(150,135){\special{em:lineto}}
\put(155,130){\special{em:lineto}}
\put(145,170){\special{em:moveto}}
\put(150,175){\special{em:lineto}}
\put(155,170){\special{em:lineto}}
\put(455,330){\special{em:moveto}}
\put(460,335){\special{em:lineto}}
\put(465,330){\special{em:lineto}}
\put(455,220){\special{em:moveto}}
\put(460,215){\special{em:lineto}}
\put(465,220){\special{em:lineto}}

\put(180,100){\makebox(0,0){$| \Psi \rangle$}}
\put(230,275){\makebox(0,0){$| \Phi \rangle$}}
\put(180,190){\makebox(0,0){$D_1$}}
\put(220,150){\makebox(0,0){$D_2$}}
\put(180,10){\makebox(0,0){$D_4$}}
\put(220,50){\makebox(0,0){$D_3$}}
\put(60,70){\makebox(0,0){$| \gamma \rangle_1$}}
\put(60,170){\makebox(0,0){$| \gamma \rangle_2$}}
\put(60,270){\makebox(0,0){$| \gamma \rangle_3$}}
\put(125,70){\makebox(0,0){$| \tilde{g}_1 \rangle_1$}}
\put(125,170){\makebox(0,0){$| \tilde{g}_1 \rangle_2$}}
\put(125,275){\makebox(0,0){$| \tilde{g}_1 \rangle_3$}}
\put(60,320){\makebox(0,0){$| v \rangle$}}
\put(90,200){\makebox(0,0){$F$}}
\put(490,350){\makebox(0,0){$D_5$}}
\put(540,300){\makebox(0,0){$D_6$}}
\put(490,200){\makebox(0,0){$D_8$}}
\put(540,250){\makebox(0,0){$D_7$}}
\put(460,250){\makebox(0,0){$M_1$}}
\put(460,300){\makebox(0,0){$M_2$}}
\put(250,220){\makebox(0,0){$U_1$}}
\put(250,330){\makebox(0,0){$U_2$}}
\put(315,275){\makebox(0,0){$U_{path}$}}
\put(380,220){\makebox(0,0){$V_1$}}
\put(380,330){\makebox(0,0){$V_2$}}
\put(190,330){\makebox(0,0){$\sigma$}}
\end{picture}

\caption{The complete setup for implementing all possible POVMs of two-photon polarization states. The state $| \Psi \rangle$ to be measured (of the form given in eq. \ref{psi}) is teleported onto the Hilbert space of the path and polarization states of photon 3 of the state $| \tilde{g}_1 \rangle$ (created by passing photons 2 and 3 of state $| \gamma \rangle$ as well as a vacuum ancilla $| v \rangle$ through a Fredkin (F) gate). The teleportation is successful if the detectors $D_1$ to $D_4$ all register one photon and the operator $\sigma_{path,z} \sigma_{pol,z} \sigma_{path,x} \sigma_{pol,x}$ (denoted by $\sigma$) is applied to the path and polarization state space of photon 3. The resulting state is $| \Phi \rangle$ which is then operated on by the SU(4) operator implemented by the unitary operators $U_1$, $U_2$, $V_1$, $V_2$ and $U_{path}$. A module of the type shown in Fig. \ref{module} with $U^s = V^s_1 = V^s_2 = I$ is placed in each of the path state arms ($M_1$ and $M_2$) and all four possible outcomes are monitored by detectors $D_5$ to $D_8$.}\label{complete}
\end{figure}

The complete setup for implementation of every mathematically possible POVM on a two-photon polarization state is shown in Fig. \ref{complete}.

\section{Example}

Our setup allows us to perform a simultaneous measurement of all four Bell states for a two photon polarization state. This task is impossible when using only linear optics on a conventional two-photon Bell state, but has been shown to be possible for Bell states in path-polarization space \cite{hardypop}. The parameters in $\tilde{U}$ and the diagonal matrix are: $\alpha = {\pi \over 4}$, $U_1 = U_2 = V_1 = I$ and $V_2 = A$, $\zeta = \xi = 0$, $\theta_1 = 0$, $\phi_1 = {\pi \over 2}$, $\theta_2 = 0$ and $\phi_2 = {\pi \over 2}$, where:

\[
A = \left(
\begin{array}{cc}
0 & 1 \cr -1 & 0 \cr
\end{array}\right)
\]
This gives rise to:
\begin{eqnarray}
F_1 &=& {1 \over \sqrt{2}} \left(
\begin{array}{cc}
 {\bf 1} & - {\bf 1}\cr
A^\dagger & A^\dagger \cr
\end{array}
\right)
\left(\begin{array}{cccc}
1 & 0 & 0 & 0 \cr
0 & 0 & 0 & 0 \cr
0 & 0 & 0 & 0 \cr
0 & 0 & 0 & 0 \cr
\end{array}\right)
{1 \over \sqrt{2}} \left(
\begin{array}{cc}
 {\bf 1} & A\cr
-{\bf 1} & A \cr
\end{array}
\right)
\cr\cr
&=& {1 \over 2}
\left(\begin{array}{cccc}
1 & 0 & 0 & 1 \cr
0 & 0 & 0 & 0 \cr
0 & 0 & 0 & 0 \cr
1 & 0 & 0 & 1 \cr
\end{array}\right) = | \Phi_+ \rangle \langle \Phi_+ |
\cr
\end{eqnarray}
and similarly:
\begin{eqnarray}
F_2 &=& | \Psi_- \rangle \langle \Psi_- |\cr
F_3 &=& | \Phi_- \rangle \langle \Phi_- |\cr
F_4 &=& | \Psi_+ \rangle \langle \Psi_+ |\cr
\end{eqnarray}
where $| \Psi_\pm \rangle = {1 \over \sqrt{2}} (| HV \rangle \pm | VH \rangle)$ and $| \Phi_\pm \rangle = {1 \over \sqrt{2}} (| HH \rangle \pm | VV \rangle)$ are the four Bell states. Thus, $F_1$ to $F_4$ are the four Bell-state projectors, as required. 

\section{Conclusion}
 
We have presented an implementation of all possible POVMs of two-photon polarization states which can be realized using existing technologies. As an example we list the settings for a simultaneous Bell State measurement. The crucial step in our setup is the teleportation of the bipartite photon polarization state to the Hilbert space of path and polarization states of one photon. This allows us to overcome many of the restrictions usually in place when manipulating and measuring bipartite systems using linear optics. 

Sebastian Ahnert was supported by the Howard Research Studentship of Sidney Sussex College, Cambridge.


\begin{thebibliography}{99}
\bibitem{NielsenChuang}
M. A. Nielsen and I. L. Chuang,
{\em Quantum Computation and Quantum Information} Cambridge University Press (2000)
\bibitem{Peres}
A. Peres, 
Phys. Lett. A {bf 128}, 19 (1988) 
\bibitem{Calsamiglia}
J. Calsamiglia, 
Phys. Rev. A {\bf 65}, 030301(R) (2002)
\bibitem{us}
S. E. Ahnert and M. C. Payne, 
Phys. Rev. A {\bf 71} 012330 (2005)
\bibitem{bergou}
M. Mohseni, A. M. Steinberg, and J. A. Bergou,
Phys. Rev. Lett. {\bf 93}, 200403 (2004) 
\bibitem{peres}
A. Peres,
{\em Quantum Theory: Concepts and Methods} (Kluwer, Dordrecht, 1993)
\bibitem{ekert}
A. K. Ekert, B. Huttner, G. M. Palma, and A. Peres, 
Phys. Rev. A {\bf 50}, 1047 (1994)
\bibitem{brandt97}
H. E. Brandt, J. M. Myers, and S. J. Lomonaco, Jr.,
Phys. Rev. A. {\bf 56} 4456 (1997); Erratum {\bf 58}, 2617 (1998)
\bibitem{brandt99}
H. E. Brandt
Am. J. Phys. {\bf 67}, 434 (1999)
\bibitem{kim}
Y.-H. Kim, S. P. Kulik, and Y. Shih 
Phys. Rev. Lett. {\bf 86}, 1370 (2001)
\bibitem{hardypop}
D. Boschi, S. Branca, F. De Martini, L. Hardy and S. Popescu
Phys. Rev. Lett. {\bf 80}, 001121 (1998) 
\bibitem{kwiat}
P. G. Kwiat and H. Weinfurter,
Phys. Rev. A {\bf 58}, R2623 (1998)
\bibitem{cinelli}
C. Cinelli, M. Barbieri, F. De Martini, and P. Mataloni,
preprint archive: quant-ph/0406148
\bibitem{kwiat2}
J. T. Barreiro, N. K. Langford, N. A. Peters, and P. G. Kwiat,
preprint archive: quant-ph/0507128 
\bibitem{bennett}
C. H. Bennett, G. Brassard, C. Cr\'epeau, R. Jozsa, A. Peres, and W. K. Wootters, 
Phys. Rev. Lett. {\bf 70}, 1895 (1993)
\bibitem{vaidman}
L. Vaidman,
Phys. Rev. A {\bf 49}, 1473 (1994)
\bibitem{braunstein}
S. L. Braunstein and H. J. Kimble,
Phys. Rev. Lett. {\bf 80}, 869 (1998)
\bibitem{stenholm}
S. Stenholm and P. J. Bardroff,
Phys. Rev. A {\bf 58}, 4373 (1998)
\bibitem{lee}
J. Lee, H. Min, and S. D. Oh, 
Phys. Rev. A {\bf 66}, 052318 (2002)
\bibitem{rigolin}
G. Rigolin,
Phys. Rev. A 71, 032303 (2005)
\bibitem{cola}
M. M. Cola and M. G. A. Paris, 
Phys. Lett. A {\bf 337}, 10 (2005)
\bibitem{milburnfredkin}
G. J. Milburn,
Phys. Rev. Lett. {\bf 62}, 2124 (1989)
\bibitem{fredkin2}
H. F. Chau and F. Wilczek,
Phys. Rev. Lett. {\bf 75}, 748 (1995)
\bibitem{fredkin3}
C. C. Gerry and R. A. Campos,
Phys. Rev. A {\bf 64}, 063814 (2001)
\bibitem{bibkraus}
K. Kraus, 
{\em Lecture Notes: States, Effects and Operations} (Springer, New York, 1983).
\end{thebibliography}
\end{document}